\documentclass[aps,prd,twocolumn,superscriptaddress,showpacs,preprintnumbers]{revtex4}
\usepackage{graphicx} 
\usepackage{dcolumn}  
\usepackage{amsmath,amssymb,amsthm}

\DeclareUnicodeCharacter{2212}{\textendash}

\usepackage[colorlinks=true]{hyperref}

\begin{document}
\vspace*{-3\baselineskip}
\resizebox{!}{2.5cm}{\includegraphics{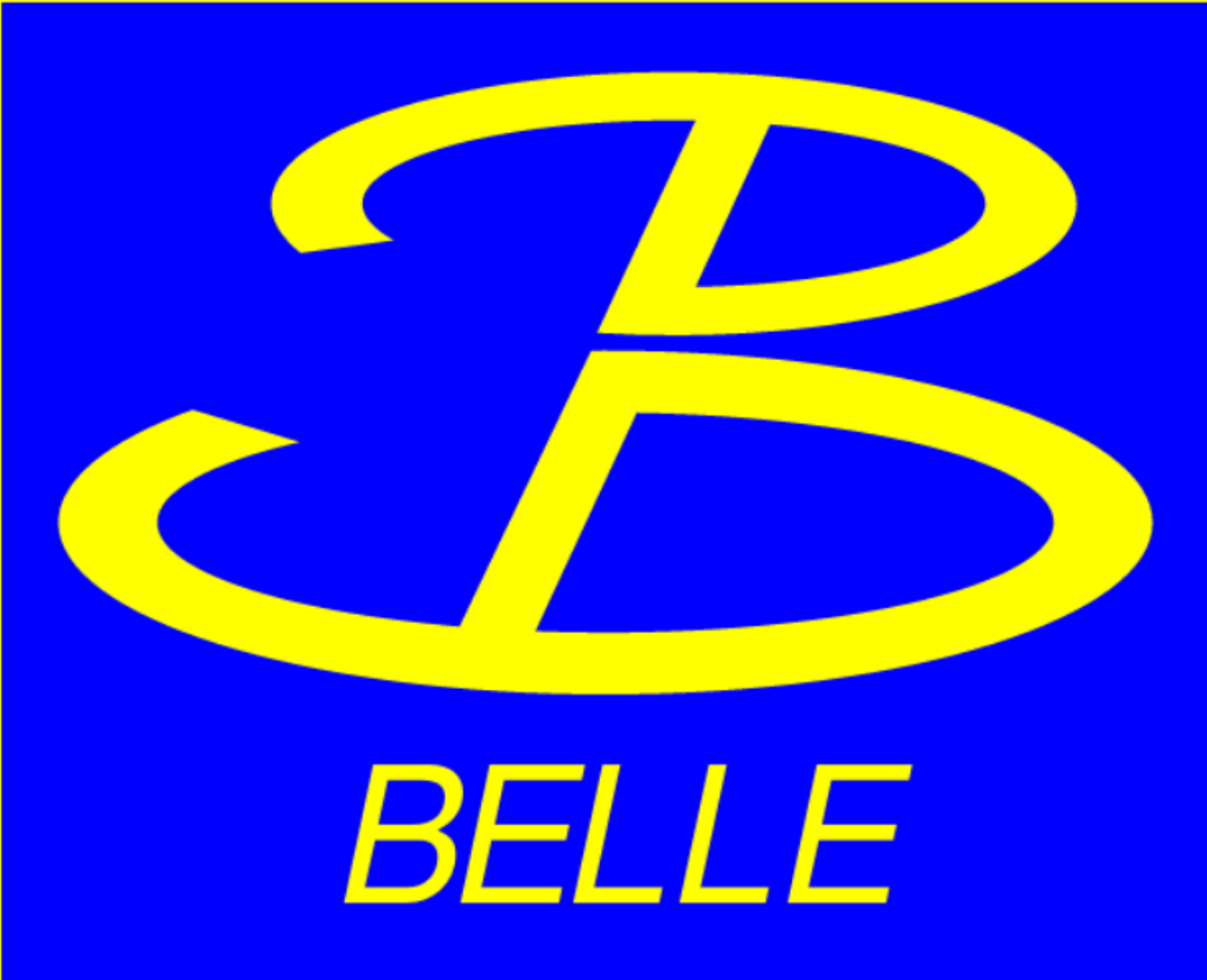}}

\preprint{\vbox{ \hbox{Belle Preprint 2021-20 v3}
    \hbox{KEK Preprint 2021-24 v3}
    \hbox{2022/06/29}
  }}

\title{ \quad\\[0.5cm] Search for $Z' \rightarrow \mu^+ \mu^-$ in the $L_\mu -
  L_\tau$ gauge-symmetric model at Belle }

\noaffiliation
\affiliation{Department of Physics, University of the Basque Country UPV/EHU, 48080 Bilbao}
\affiliation{University of Bonn, 53115 Bonn}
\affiliation{Brookhaven National Laboratory, Upton, New York 11973}
\affiliation{Budker Institute of Nuclear Physics SB RAS, Novosibirsk 630090}
\affiliation{Faculty of Mathematics and Physics, Charles University, 121 16 Prague}
\affiliation{Chonnam National University, Gwangju 61186}
\affiliation{University of Cincinnati, Cincinnati, Ohio 45221}
\affiliation{Deutsches Elektronen--Synchrotron, 22607 Hamburg}
\affiliation{Duke University, Durham, North Carolina 27708}
\affiliation{University of Florida, Gainesville, Florida 32611}
\affiliation{Department of Physics, Fu Jen Catholic University, Taipei 24205}
\affiliation{Key Laboratory of Nuclear Physics and Ion-beam Application (MOE) and Institute of Modern Physics, Fudan University, Shanghai 200443}
\affiliation{Justus-Liebig-Universit\"at Gie\ss{}en, 35392 Gie\ss{}en}
\affiliation{Gifu University, Gifu 501-1193}
\affiliation{SOKENDAI (The Graduate University for Advanced Studies), Hayama 240-0193}
\affiliation{Gyeongsang National University, Jinju 52828}
\affiliation{Department of Physics and Institute of Natural Sciences, Hanyang University, Seoul 04763}
\affiliation{University of Hawaii, Honolulu, Hawaii 96822}
\affiliation{High Energy Accelerator Research Organization (KEK), Tsukuba 305-0801}
\affiliation{J-PARC Branch, KEK Theory Center, High Energy Accelerator Research Organization (KEK), Tsukuba 305-0801}
\affiliation{Higher School of Economics (HSE), Moscow 101000}
\affiliation{Forschungszentrum J\"{u}lich, 52425 J\"{u}lich}
\affiliation{IKERBASQUE, Basque Foundation for Science, 48013 Bilbao}
\affiliation{Indian Institute of Science Education and Research Mohali, SAS Nagar, 140306}
\affiliation{Indian Institute of Technology Bhubaneswar, Satya Nagar 751007}
\affiliation{Indian Institute of Technology Guwahati, Assam 781039}
\affiliation{Indian Institute of Technology Hyderabad, Telangana 502285}
\affiliation{Indian Institute of Technology Madras, Chennai 600036}
\affiliation{Indiana University, Bloomington, Indiana 47408}
\affiliation{Institute of High Energy Physics, Chinese Academy of Sciences, Beijing 100049}
\affiliation{Institute of High Energy Physics, Vienna 1050}
\affiliation{Institute for High Energy Physics, Protvino 142281}
\affiliation{INFN - Sezione di Napoli, 80126 Napoli}
\affiliation{INFN - Sezione di Torino, 10125 Torino}
\affiliation{Advanced Science Research Center, Japan Atomic Energy Agency, Naka 319-1195}
\affiliation{J. Stefan Institute, 1000 Ljubljana}
\affiliation{Institut f\"ur Experimentelle Teilchenphysik, Karlsruher Institut f\"ur Technologie, 76131 Karlsruhe}
\affiliation{Kavli Institute for the Physics and Mathematics of the Universe (WPI), University of Tokyo, Kashiwa 277-8583}
\affiliation{Kitasato University, Sagamihara 252-0373}
\affiliation{Korea Institute of Science and Technology Information, Daejeon 34141}
\affiliation{Korea University, Seoul 02841}
\affiliation{Kyoto Sangyo University, Kyoto 603-8555}
\affiliation{Kyungpook National University, Daegu 41566}
\affiliation{Universit\'{e} Paris-Saclay, CNRS/IN2P3, IJCLab, 91405 Orsay}
\affiliation{P.N. Lebedev Physical Institute of the Russian Academy of Sciences, Moscow 119991}
\affiliation{Faculty of Mathematics and Physics, University of Ljubljana, 1000 Ljubljana}
\affiliation{Ludwig Maximilians University, 80539 Munich}
\affiliation{Luther College, Decorah, Iowa 52101}
\affiliation{Malaviya National Institute of Technology Jaipur, Jaipur 302017}
\affiliation{Faculty of Chemistry and Chemical Engineering, University of Maribor, 2000 Maribor, Slovenia}
\affiliation{Max-Planck-Institut f\"ur Physik, 80805 M\"unchen}
\affiliation{School of Physics, University of Melbourne, Victoria 3010}
\affiliation{University of Mississippi, University, Mississippi 38677}
\affiliation{University of Miyazaki, Miyazaki 889-2192}
\affiliation{Moscow Physical Engineering Institute, Moscow 115409}
\affiliation{Graduate School of Science, Nagoya University, Nagoya 464-8602}
\affiliation{Kobayashi-Maskawa Institute, Nagoya University, Nagoya 464-8602}
\affiliation{Universit\`{a} di Napoli Federico II, 80126 Napoli}
\affiliation{Nara Women's University, Nara 630-8506}
\affiliation{National Central University, Chung-li 32054}
\affiliation{National United University, Miao Li 36003}
\affiliation{Department of Physics, National Taiwan University, Taipei 10617}
\affiliation{H. Niewodniczanski Institute of Nuclear Physics, Krakow 31-342}
\affiliation{Nippon Dental University, Niigata 951-8580}
\affiliation{Niigata University, Niigata 950-2181}
\affiliation{University of Nova Gorica, 5000 Nova Gorica}
\affiliation{Novosibirsk State University, Novosibirsk 630090}
\affiliation{Osaka City University, Osaka 558-8585}
\affiliation{Pacific Northwest National Laboratory, Richland, Washington 99352}
\affiliation{Panjab University, Chandigarh 160014}
\affiliation{Peking University, Beijing 100871}
\affiliation{University of Pittsburgh, Pittsburgh, Pennsylvania 15260}
\affiliation{Research Center for Nuclear Physics, Osaka University, Osaka 567-0047}
\affiliation{Meson Science Laboratory, Cluster for Pioneering Research, RIKEN, Saitama 351-0198}
\affiliation{Department of Modern Physics and State Key Laboratory of Particle Detection and Electronics, University of Science and Technology of China, Hefei 230026}
\affiliation{Seoul National University, Seoul 08826}
\affiliation{Showa Pharmaceutical University, Tokyo 194-8543}
\affiliation{Soochow University, Suzhou 215006}
\affiliation{Soongsil University, Seoul 06978}
\affiliation{Sungkyunkwan University, Suwon 16419}
\affiliation{School of Physics, University of Sydney, New South Wales 2006}
\affiliation{Department of Physics, Faculty of Science, University of Tabuk, Tabuk 71451}
\affiliation{Tata Institute of Fundamental Research, Mumbai 400005}
\affiliation{Department of Physics, Technische Universit\"at M\"unchen, 85748 Garching}
\affiliation{Thomas Jefferson National Accelerator Facility, Newport News VA 23606}
\affiliation{Department of Physics, Tohoku University, Sendai 980-8578}
\affiliation{Earthquake Research Institute, University of Tokyo, Tokyo 113-0032}
\affiliation{Department of Physics, University of Tokyo, Tokyo 113-0033}
\affiliation{Tokyo Institute of Technology, Tokyo 152-8550}
\affiliation{Tokyo Metropolitan University, Tokyo 192-0397}
\affiliation{Utkal University, Bhubaneswar 751004}
\affiliation{Virginia Polytechnic Institute and State University, Blacksburg, Virginia 24061}
\affiliation{Wayne State University, Detroit, Michigan 48202}
\affiliation{Yamagata University, Yamagata 990-8560}
\affiliation{Yonsei University, Seoul 03722}

  \author{T.~Czank}\affiliation{Kavli Institute for the Physics and Mathematics of the Universe (WPI), University of Tokyo, Kashiwa 277-8583} 
  \author{I.~Jaegle}\affiliation{University of Florida, Gainesville, Florida 32611}\affiliation{Thomas Jefferson National Accelerator Facility, Newport News VA 23606} 
  \author{A.~Ishikawa}\affiliation{High Energy Accelerator Research Organization (KEK), Tsukuba 305-0801}\affiliation{SOKENDAI (The Graduate University for Advanced Studies), Hayama 240-0193} 

\author{I.~Adachi}\affiliation{High Energy Accelerator Research Organization (KEK), Tsukuba 305-0801}\affiliation{SOKENDAI (The Graduate University for Advanced Studies), Hayama 240-0193} 
  \author{K.~Adamczyk}\affiliation{H. Niewodniczanski Institute of Nuclear Physics, Krakow 31-342} 
  \author{H.~Aihara}\affiliation{Department of Physics, University of Tokyo, Tokyo 113-0033} 
  \author{D.~M.~Asner}\affiliation{Brookhaven National Laboratory, Upton, New York 11973} 
  \author{T.~Aushev}\affiliation{Higher School of Economics (HSE), Moscow 101000} 
  \author{R.~Ayad}\affiliation{Department of Physics, Faculty of Science, University of Tabuk, Tabuk 71451} 
  \author{V.~Babu}\affiliation{Deutsches Elektronen--Synchrotron, 22607 Hamburg} 
% ...
\collaboration{The Belle Collaboration}

\begin{abstract}
  We search for a new gauge boson $Z'$ that couples only to heavy leptons and
  their corresponding neutrinos in the process $e^{+} e^{-} \rightarrow Z'(\rightarrow
  \mu^{+}\mu^{-}) \mu^{+}\mu^{-}$, using a 643 fb$^{-1}$ data sample collected by the
  Belle experiment at or near the $\Upsilon(1S,2S,3S,4S,5S)$ resonances at the KEKB
  collider. While previous searches for $Z'$ did a data-based estimation of the
  initial state radiation effect, our search for the $Z'$ is the first to include
  effects due to initial state radiation in the signal simulated samples used in estimating the detection
  efficiency. No signal is observed in the $Z'$ mass range of 0.212 \--- 10 GeV$/c^2$
  and we set an upper limit on the coupling strength, $g'$, constraining the
  possible $Z'$ contribution to the anomalous magnetic dipole moment of the muon. 
\end{abstract}


\maketitle

\section{Introduction}
\label{sec-basic}

The lack of evidence for a Weakly Interacting Massive Particle by underground
experiments~\cite{UG1, UG2}, and the absence of supersymmetric particle signals at the LHC~\cite{LHC1,LHC2,LHC3}, suggest that dark matter might be composite and/or light. This gives
rise to dark sector models~\cite{Pospelov2007,Arkani2008,Chun2008,Cheung2009,Katz2009,Morrissey2009,Goodsell2009,Baumgart2009,Nomura2008,Alves2009,Jaeckel2010}
that introduce a zoology of dark particles which do not interact directly via Standard
Model (SM) forces, but can interact by dark sector forces via new mediators and
therefore only indirectly with (SM) particles, and could have masses between
1 MeV/${\it c}^2$ and 10 GeV/${\it c}^2$.

Discrepancies observed at low-energy measurements~\cite{gm2_2006, Barger:2010aj} have fueled new precision studies.
Within this context, the anomalous
magnetic moment of the muon, $(g-2)_\mu$, is one of the most precisely
measured quantities in particle physics, where the difference between the
experimental value and the SM prediction~\cite{AOYAMA20201} is about 4.2$\sigma$~\cite{gm2_precision}.
This discrepancy might be a sign of new physics and has led to a variety of
attempts to create physics models involving the leptonic sector of the SM~\cite{firstzp, zpmodel,
  neutrinotrident, brianpap}.

These attempts include the set of SM extensions which add a new $U(1)$ gauge
boson ($Z'$) coupled to the difference between lepton family numbers, $L_i$ where
$i=e, \mu$ and $\tau$~\cite{zpmodel}.
The electron number differences have been well constrained by measurements performed at $e^+
e^-$ colliders~\cite{Essig:2009nc, BaBar:2014zli} and will not be discussed
here.
In this study we present a search for the gauge boson coupled to the $L_\mu
- L_\tau$ difference.

The partial widths for the $Z'$ decay to leptons~\cite{brianprivate, darkpho-hepcol} are given by: 

\begin{align}\label{eq:zplept}
  \Gamma(Z'\rightarrow \ell^+ \ell^- ) & =  \frac{(g')^2 M_{Z'}}{12 \pi}  \left( 1 + \frac{2M^{2}_{\ell}}{M^{2}_{Z'}}\right)  \nonumber \\
                                       & \qquad \sqrt{1-\frac{4M^{2}_{\ell}}{M^{2}_{Z'}}}\theta(M_{Z'} - 2M_{\ell})
\end{align}
where $g'$ is the $L_\mu - L_\tau$ coupling strength, and $\theta(M_{Z'} -
2M_{\ell})$ is a step function, and

\begin{align}\label{eq:zpneutri}
\Gamma(Z'\rightarrow \nu_{\ell} \bar{\nu}_{\ell} ) = \frac{(g')^2 M_{Z'}}{24 \pi}.
\end{align}
For $M_{Z'} \gg M_{\ell}$ the branching fraction to one neutrino flavor is half
of that to a charged lepton. This is due to the fact that the $Z'$ boson only couples to
left-handed neutrinos, but couples to both left- and right-handed charged leptons.

The visible branching fraction to muons is:

\begin{align}
\label{eq:zpmumu}
  \mathcal{B}(Z'\rightarrow \mu^+ \mu^-) &=& \frac{\Gamma(Z'\rightarrow \mu^+ \mu^-)}{\sum_{\ell}^{\mu,\tau} (\Gamma(Z'\rightarrow \nu_{\ell} \bar{\nu}_{\ell}) + \Gamma(Z'\rightarrow \ell^+ \ell^-))},
 \end{align}
which is identical to $\mathcal{B}(Z'\rightarrow \tau \tau)$ except
for the replacement of the decay width with the appropriate decay channel.

\begin{figure}[h!]
  \centering
  \includegraphics[scale=0.3]{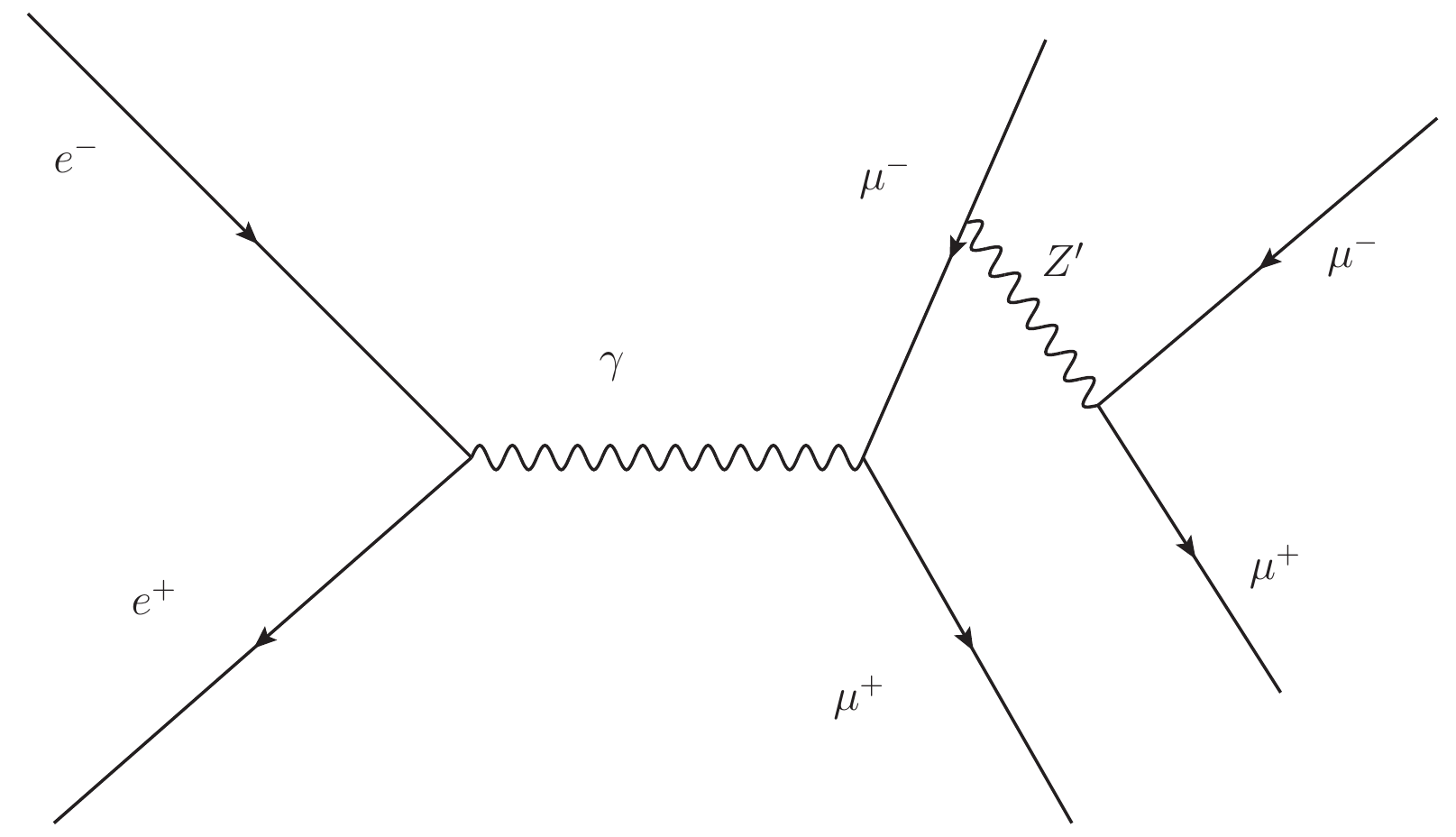}
  \caption{\label{fig::bellezpfeyn} Feynman diagram for the main production
    channel of the $Z'$ in $e^+ e^-$ colliders.}
\end{figure}

We search for the $Z'$ of an $L_\mu -L_\tau$ model via the decay $Z'
\rightarrow \mu^+ \mu^-$. In this model, the $Z'$ only couples to the second and
third generation of leptons
($\mu, \tau$) and their neutrinos.
We search for four-muon events in the reaction depicted in
Fig.~\ref{fig::bellezpfeyn}, in which the $e^+ e^- \rightarrow  \mu^+
\mu^-$ process is followed by $Z'$ radiation from a muon, and then, the $Z'$ decays to $\mu^+ \mu^-$.  

In addition to its possible contribution to the $(g-2)_\mu$ anomaly, the effects
of a $Z'$ have been searched for in other scenarios. It could be a source of an increase in
neutrino trident production $\nu_\mu N \rightarrow N \nu_\mu \mu^+ \mu^-$
~\cite{neutrinotrident}. No increase has been observed, and a limit was set for the
$Z'$ parameter space.
It could also work as an indirect channel to sterile neutrino dark matter ~\cite{brianpap}, and could
provide predictions for the neutrino mass-mixing matrix ~\cite{asai2018,
  asai2017, ma2001}.

Recently, the Belle II collaboration published the search result with $Z' \rightarrow
\nu \nu$ decay using a 276 pb$^{-1}$ luminosity data \cite{b2invzp}. No $Z'$
signature was found so an upper limit of the parameter space of this decay mode
was set.
Previously BABAR searched for the $Z'$ with $e^+ e^- \rightarrow
Z'(\rightarrow \mu^+ \mu^-) \mu^+ \mu^-$ using a 514 fb$^{-1}$ luminosity data
and since no $Z'$ signature was found the most stringent upper limits as a function of $Z'$
mass~\cite{babar2016zp} was set.
In this paper, we present a search for the same $Z'$ model in the full available Belle data
sample.

\section{Experimental Setup}

The search for $e^+ e^- \rightarrow \mu^+ \mu^- Z' (\rightarrow \mu^+ \mu^-)$
is performed using the following luminosities: 33 fb$^{-1}$ taken at
the $\Upsilon(1S)$ and $\Upsilon(2S)$ resonances, 2 fb$^{-1}$ at the $\Upsilon(3S)$ resonance, 484 fb$^{-1}$
at the $\Upsilon(4S)$ resonance, 93 fb$^{-1}$ at the $\Upsilon(5S)$ resonance,
and 67 fb$^{-1}$ taken 60 MeV below the $\Upsilon(4S)$ resonance,
totaling 679 fb$^{-1}$ collected by the Belle detector
~\cite{belle_0, belle_1} at the KEKB collider ~\cite{kekb_0, kekb_1}. A 36 fb$^{-1}$ subset of
the $\Upsilon(4S)$ sample, the validation sample, is used to verify the
selection criteria and then discarded from the analysis.

The Belle detector surrounds the interaction
point of KEKB. It is a large-solid-angle magnetic spectrometer consisting of
a silicon vertex detector, a 50-layer central drift chamber (CDC), an array of aerogel threshold
Cherenkov counters, a barrel-like arrangement of
time-of-flight scintillation counters, and an electromagnetic calorimeter (ECL) comprised of CsI(TI) crystals
located inside a superconducting solenoid coil that provides
a 1.5 T magnetic field. An iron flux return located outside
of the coil is instrumented with resistive plate chambers to detect $K_{L}^{0}$ mesons and
identify muons (KLM).
The signal events are guaranteed to
pass the trigger with nearly full efficiency because the muonic $Z'$ decay
topology features more than three charged tracks. In addition, the large radius
of the CDC (880 mm) ~\cite{bfactories} allows an excellent mass resolution and muon
detection efficiency in Belle.

\section{Selection Criteria}

The selection is optimized based on the validation sample as well as a Monte Carlo (MC) simulation done in two steps.
First, signal events are generated for different $Z'$ mass hypotheses using
\texttt{WHIZARD}~\cite{madgraph}, which takes into account the Initial State
Radiation (ISR) as well as the Final State Radiation (FSR) at the
$\Upsilon(4S)$ center-of-mass energy. \texttt{WHIZARD} also has an option to
generate events without radiative corrections. Then, the detector response to these
events is simulated using \texttt{GEANT3}~\cite{geant3}.
There were 54 mass hypotheses generated for each of the $Z'$ MC samples from
$m_{Z'} = 212$ MeV$/c^2$ to $m_{Z'} = 1.015$ GeV$/c^2$ in 100 MeV$/c^2$ steps, and
subsequently in 200 MeV$/c^2$ steps up to $m_{Z'} = 10.00$ GeV$/c^2$.
The change in the steps is due to the behavior of the detection efficiency observed
in the analysis.

The irreducible background, $e^+ e^- \rightarrow \mu^+ \mu^- \mu^+ \mu^-$
is studied with an MC sample corresponding to a luminosity of 336
fb$^{-1}$ generated with \texttt{Diag36}~\cite{diag36} at
$\Upsilon(4S)$ center-of-mass energy, \texttt{Diag36} generates events without
ISR corrections (non-ISR). There is no event generator available for the
QED 4-lepton final state with radiative correction. In addition, other leptonic and hadronic
background sources, such as $e^+ e^- \rightarrow e^+ e^- e^+ e^-$ and $e^+ e^-
\rightarrow \pi^+ \pi^- J/\psi(\rightarrow \mu^+ \mu^-)$ were studied through MC
samples, and found to give negligible or no contributions after the application of
the selection criteria.

We select events with two pairs of oppositely charged tracks in the final state.
To ensure these tracks originate from the interaction point, their
transverse and longitudinal impact
parameters must be less than 0.2 and 1.5 cm, respectively.
At least two tracks are required to have a muon likelihood ratio,
$\frac{\mathcal{L}_\mu}{\mathcal{L}_\mu + \mathcal{L}_K + \mathcal{L}_\pi}$,
greater than 0.1. The value of  $\mathcal{L}_\mu$ depends on the
difference between the expected and actual muon penetration of the track in the KLM, and the
distance between its KLM hits and the extrapolation of the track from the CDC. The efficiency
for a track to be identified as a muon is about 95\% for momenta between 1 to 3
GeV$/c$, and slightly lower momenta below 1 GeV$/c$.
In addition, a hadron veto is applied. The muon candidate must not have a
likelihood ratio corresponding to a pion, kaon, or a proton. This is implemented
by comparing the likelihood ratio of two particles (proton and kaon, kaon and pion, and proton and pion) as $P(i|j) = \frac{\mathcal{L}_i}{\mathcal{L}_i + \mathcal{L}_j}$, where $\mathcal{L}_i$ is the likelihood product from
three detectors (ACC, TOF, and CDC). A pion is defined as $P(K|\pi)<0.4$ and $P(p|\pi)<0.4 $. A kaon is defined as $P(p|K)<0.4$ and $P(K|i)>0.6$. A proton is defined as $P(p|K)>0.6$ and
  $P(p|\pi)>0.6$.  

To suppress the background due to neutral particles, the sum of ECL clusters unrelated to any charged
tracks with energy greater than 30 MeV is required to be less than 200 MeV.
Additionally, the visible energy, $E_{\rm vis}$, calculated from the four muons
must be consistent with the center-of-mass energy, $E_{\rm CMS}$, so that $|
E_{\rm CMS} - E_{\rm vis}| < 500$~MeV.

A kinematic fit based on the least square method for the final state is carried out under the constraint that the
four-momentum of the final state be compatible with the initial $e^+ e^-$
system. The chi-squared is minimized by Lagrange multipliers,
they issue a set of non-linear equations that are solved using the multi-dimensional
Newton-Raphson method. As a result, the reconstructed $Z'$ mass resolution is improved.

As there are four possible combinations of oppositely charged muons in the
final state, all four possible combinations correspond to four $Z'$ candidates counted
per event.
To improve the sensitivity in the low $Z'$ mass region, we introduce a reduced
mass, defined as $m_R = \sqrt{m^{2}_{\mu^+ \mu^-} -
 4m_{\mu, \mathrm{PDG}}^2}$, where $m_{\mu \mu}$ is the invariant masses of
$Z'$ candidate and $m_{\mu, \mathrm{PDG}}^2$ is the muon nominal mass~\cite{Zyla:2020zbs}. The $m_R$ distribution is smoother
than the invariant mass distribution around the $Z'$ mass close to the dimuon threshold.

\begin{figure}[h!]
\centering
\includegraphics[scale=0.4]{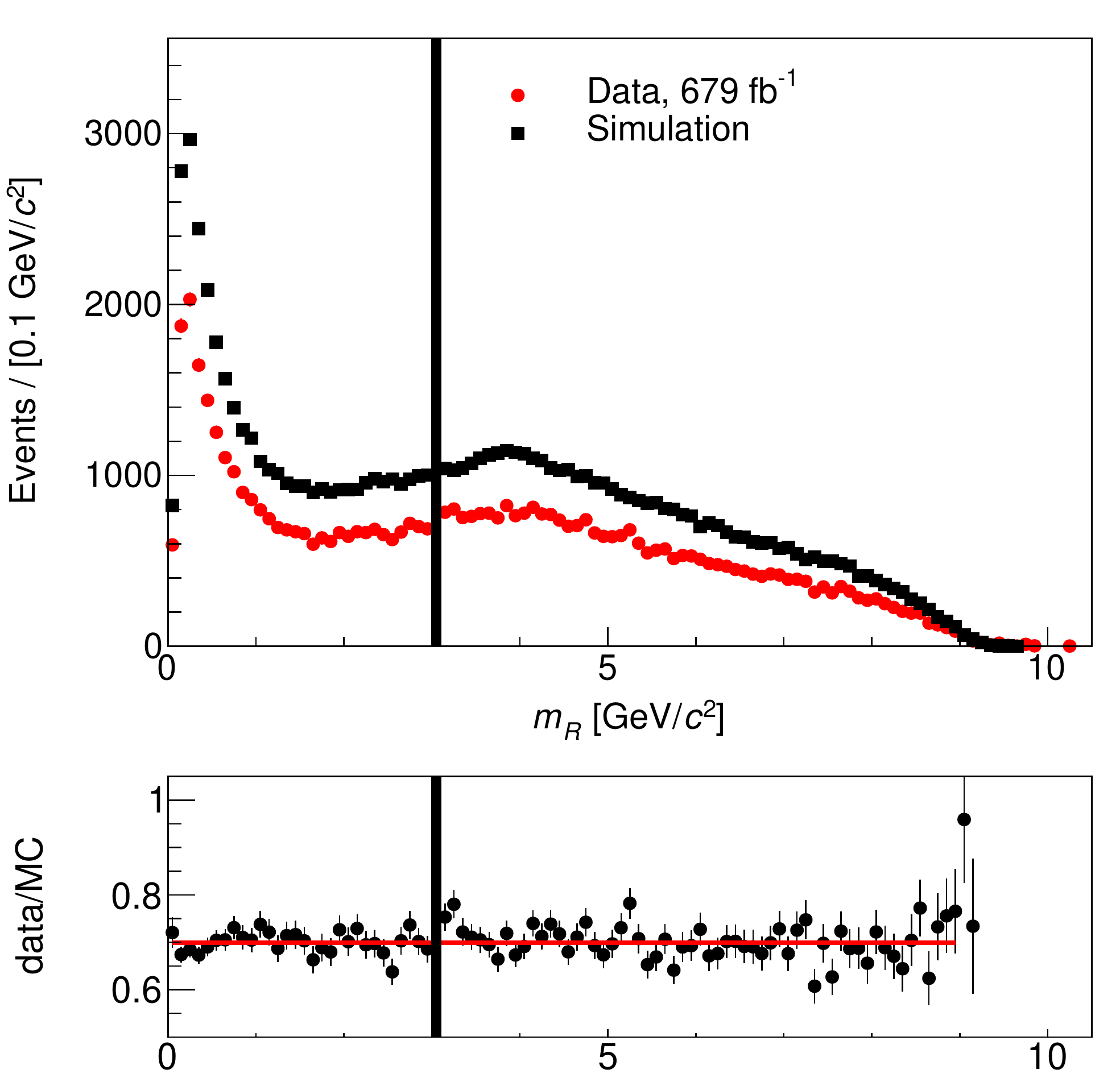}
\caption{\label{fig::redmasseff}
  (Top) Reduced mass, $m_R$, distributions. Red points represent the data after
  all selection criteria is applied. Black squares represent the non-ISR MC
  expectation for the $e^+ e^- \rightarrow \mu^+ \mu^-
  \mu^+\mu^-$ (\texttt{Diag36})~\cite{diag36} scaled to the data luminosity.
  (Bottom) The ratio between data and the non-ISR $e^+ e^- \rightarrow \mu^+
  \mu^- \mu^+ \mu^-$ MC expectation. The red line represents a fit of a 1st order
  polynomial where its constant term is $0.700 \pm 0.003$ and the slope term is negligible.              
  (Both) Black shaded region at 3.1 GeV$/c^2$ represents the $J/\psi$ region
  which is not used in this analysis.
}
\end{figure}

The $Z'$ reduced mass distributions for data and MC are compared in Fig.~\ref{fig::redmasseff}.
Although the normalization of the data is almost 70\% of the background level,
determined by a fit to a constant probability density function (pdf) as shown in
Fig.~\ref{fig::redmasseff} (bottom). This difference
arises due to the ISR effect, which is not simulated in the background MC sample.

We veto the reduced mass distribution around the $J/\psi$ mass, $3.05 < m_{R} < 3.13$ GeV$/c^2$,
as its muonic decay can mimic a signal.
This was not necessary around the $\psi(2S)$ mass since the
$\psi(2S)$ decay into muons is negligible compared to the main background.

\section{Results}
\label{sec::res}
 

We perform a binned maximum-likelihood fit to the reduced mass
distribution with the range of $m_{Z'} \pm 25 \sigma_{Z'}$. The fit is repeated
9788 times with a different $Z'$ mass hypotheses in steps of
1~MeV$/c^{2}$ from 0 to 9787 MeV$/c^{2}$.
The $Z'$ resolution starts from less than 1~MeV$/c^2$ at the dimuon mass
threshold increasing until 5.5~GeV$/c^2$ where it is valued at 6~MeV$/c^2$ then
it starts decreasing until 9.5~GeV$/c^2$ where it is valued at $\sim$3~MeV$/c^2$.
The step is set around half of the width of the reduced mass distribution for
MC generated signal.

The signal $m_R$ distribution is modeled as a sum of two Crystal Ball~\cite{crystalball} functions with a common mean. The shape parameters as a function of the $m_R$ are
determined with signal MC samples while the normalization is floated in the fit. The
width is calibrated using $J/\psi \rightarrow \mu^+ \mu^-$ events in the veto
region. The background is modeled with a third-order polynomial which is the
lowest order function that can fit the $e^+ e^- \rightarrow \mu^+ \mu^- \mu^+
\mu^-$ background well. Background normalization and shape parameters are floated in
the fit.

The efficiency is determined using a fit to the MC signal samples with different
mass hypotheses. It is the result of the integration of the fit function over
$m_{Z'} \pm 3\sigma_{Z'}$, where $\sigma_{Z'}$ is the $Z'$ mass resolution. This
efficiency is interpolated between the different discrete mass hypotheses.

This procedure is done identically for non-ISR MC samples where the $m_R$
distribution is also modeled as a sum of two Crystal Ball functions with a
common mean, however, for the non-ISR MC samples the parametrization of the pdf
is different than for the ISR case.
Comparing ISR and non-ISR detection efficiencies is key to understand the gap
between data and MC background on Fig.~\ref{fig::redmasseff}.

\begin{figure}[h!]
  \includegraphics[scale=0.4]{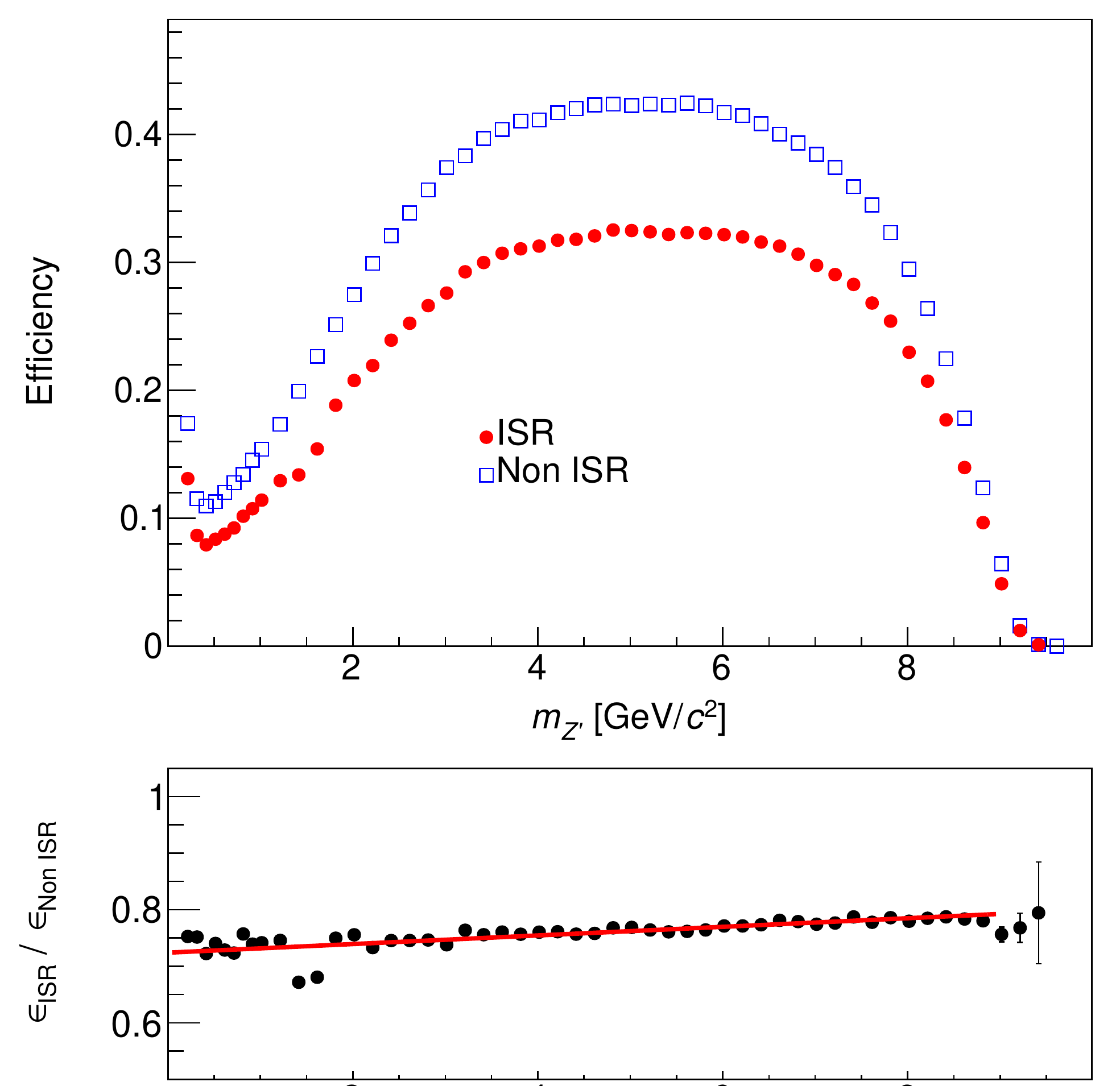}
  \caption{\label{fig::deteff}(Top) Detection efficiency as a function of the reduced
    mass. Red points represent values with ISR correction. Open blue
    squares represent values without it.
  (Bottom) Ratio between ISR and non-ISR detection efficiencies linearly fit.
  The average constant value from a linear and constant fit is $0.741 \pm 0.001$.}
\end{figure}

Fig.~\ref{fig::deteff} shows efficiencies as a function of reduced mass.

It is clear that the detection efficiency increases
with increasing $Z'$ mass up to 6 GeV/$c^2$ and then it decreases.
This behavior is due to the muon detection efficiency in the KLM, which has a threshold momentum
of 600 MeV/$c$ reaching maximum at 1 GeV/$c$ then flattens for even higher values.

Systematic uncertainties arise from luminosity, track identification, muon
identification and fitting bias.
The luminosity uncertainty is 1.4\% and is measured using Bhabha and
two-photon events. The track identification uncertainty is 0.35\%
per charged track, or 1.4\% in this analysis, and is determined by comparing the track finding
efficiency of partially and fully reconstructed $D^{*+} \rightarrow
D^0(\rightarrow K^- \pi^+) \pi^+$ decays.
A muon identification uncertainty of 1.15\% is
determined from the change in event yields while varying the muon likelihood ratio criterion from 0.1 to 0.2.
With muon likelihood cuts there is also a systematic error to be considered on
the detection efficiency calculated through MC signal samples. This error is
calculated comparing $\gamma \gamma \rightarrow \mu^+ \mu^-$ data and MC
samples. Due to the large number of these events it is possible to map the dependency
between momentum, muon likelihood ratio and error rate. Comparing our MC signal
calculated detection efficiency with and without this correction gives a 1\% difference. 
Finally, a correction from the hadron veto is implemented on the MC samples.
This correction factor is also of 1\% and it is obtained by comparing MC samples
with and without the hadron veto.

The effect of fitting bias is investigated using a bootstrap study
to check whether allowing third-order polynomial components to float in the
fit end up inducing a bias on the yield extracted. For each mass scan, this study
is done by varying the data with a Poisson distribution, varying each
individual bin of the histogram. This changed data set is then injected with a
signal of yield corresponding to a Poisson distribution of the upper limit on the number of observed events and a distribution following its pdf. This reconstructed ensemble is then
fitted in the same way as the data. The newly extracted yield, $\mathrm{N_{sig}}$, is then compared to the
true number of events injected, $\mathrm{N^{true}_{sig}}$, divided by the
uncertainty in the newly yield extracted, $\sigma_{\mathrm{N_{sig}}}$, as
$(\mathrm{N^{true}_{sig} - N_{sig}}) / \sigma_{\mathrm{N_{sig}}}$.
This procedure is repeated 1000 times for each mass scan.
We find that the extracted yield and its uncertainty are systematically
overestimated by 3$\%$ and 4$\%$, respectively. These biases are
accordingly taken into account in the $Z'$ scan and $g'$ upper limit calculation
by correcting the yield extracted and the error on the yield extracted. They
correspond to $N^{\mathrm{cor}}_{\mathrm{sig}} = N_\mathrm{sig}\left( 1 + b \right) $ and $N^{\mathrm{errcor}}_\mathrm{sig} = N^{\mathrm{err}}_\mathrm{sig}
\times b^\mathrm{err}$, where $b$ stands for bias and the variables with a
superscript $\mathrm{err}$ are related to the error on the yield.

\begin{figure}[ht]
    \centering
    \includegraphics[scale=0.4]{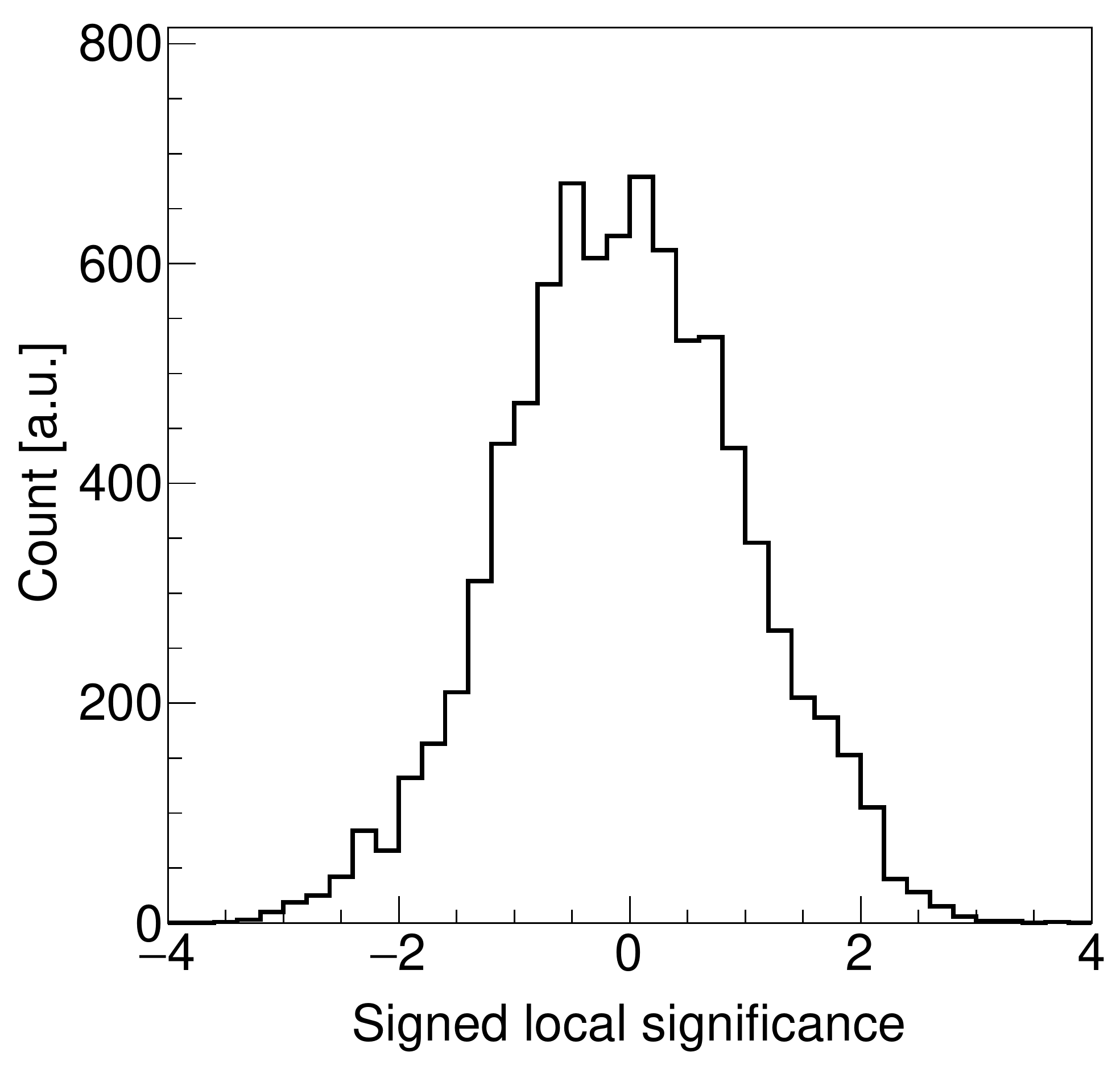}
    \caption{ Local signed significance values.
    }
    \label{fig::signif}
  \end{figure}

The significance of each possible $Z'$ candidate is evaluated as
  
  \begin{equation}
    \label{eq:signi}
   \mathcal{S}=\mathrm{sign}(N_{\mathrm{obs}})\sqrt{2\log({\mathcal{L}_{\mathrm{S+B}}/\mathcal{L}_{\mathrm{B}}})},
  \end{equation}
  where $\mathrm{sign}(N_{\mathrm{obs}})$ is the sign of the number of observed
  events and
  $\mathcal{L}_{\mathrm{S+B}} / \mathcal{L}_{\mathrm{B}}$ is the ratio between
the maximum likelihoods of the fits with signal plus background hypothesis
$(\mathcal{L}_{\mathrm{S+B}})$ and background only hypothesis $(\mathcal{L}_{\mathrm{B}})$.
  The distribution of significances is shown Fig.~\ref{fig::signif}.
  
\begin{figure}[h!]
\centering
      \includegraphics[scale=0.4]{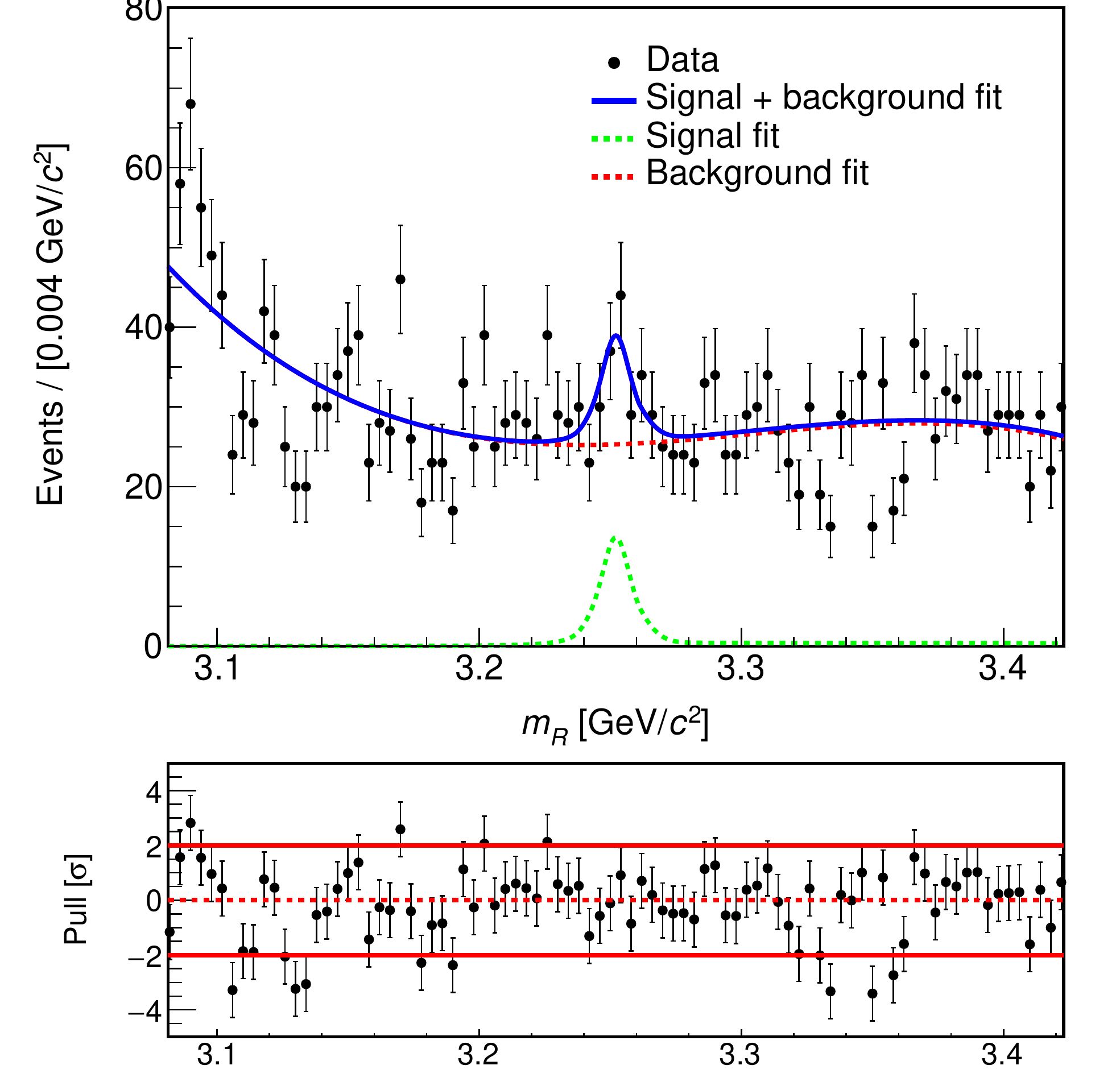}
      \caption{\label{fig::zoomfit} The solid blue curve represents a fit to the
        data of a sum of two Crystal Ball functions added to a third-order polynomial. The dashed red and green curves represent the third-order
        polynomial and the sum of two Crystal Ball functions, respectively.}
    \end{figure}

   The largest local significance observed in an excess
   (deficit) is 3.7$\sigma$ (3.5$\sigma$) around $m_{Z'} = 3.3$ GeV$/c^2$ (3.1 GeV$/c^2$), in Fig.~\ref{fig::zoomfit}. After
  incorporating the look-elsewhere-effect the global significance for the
  excess becomes 2.23$\sigma$.

  Since no fit resulted in a global significance of at least 5$\sigma$, we set upper
  limits on the coupling strength $g'$ as a function of $m_{Z'}$.
A Bayesian method~\cite{DAgostini} is used to estimate the 90\% credibility
level~(C.L.) upper limit on the number of observed signal events,
$N_\mathrm{obs}$.
A flat prior is assumed for the signal yield and two nuisance parameters are
added one for the signal yield and another for the background yield. These
nuisance parameters are two Gaussian uncertainties which correspond to the
systematic errors. For the background nuisance
parameter the statistical errors are added in quadratic sum to the systematic
errors~\cite{Moneta:2010pm}.

\begin{figure}[h!]
  \centering
  \includegraphics[scale=0.4]{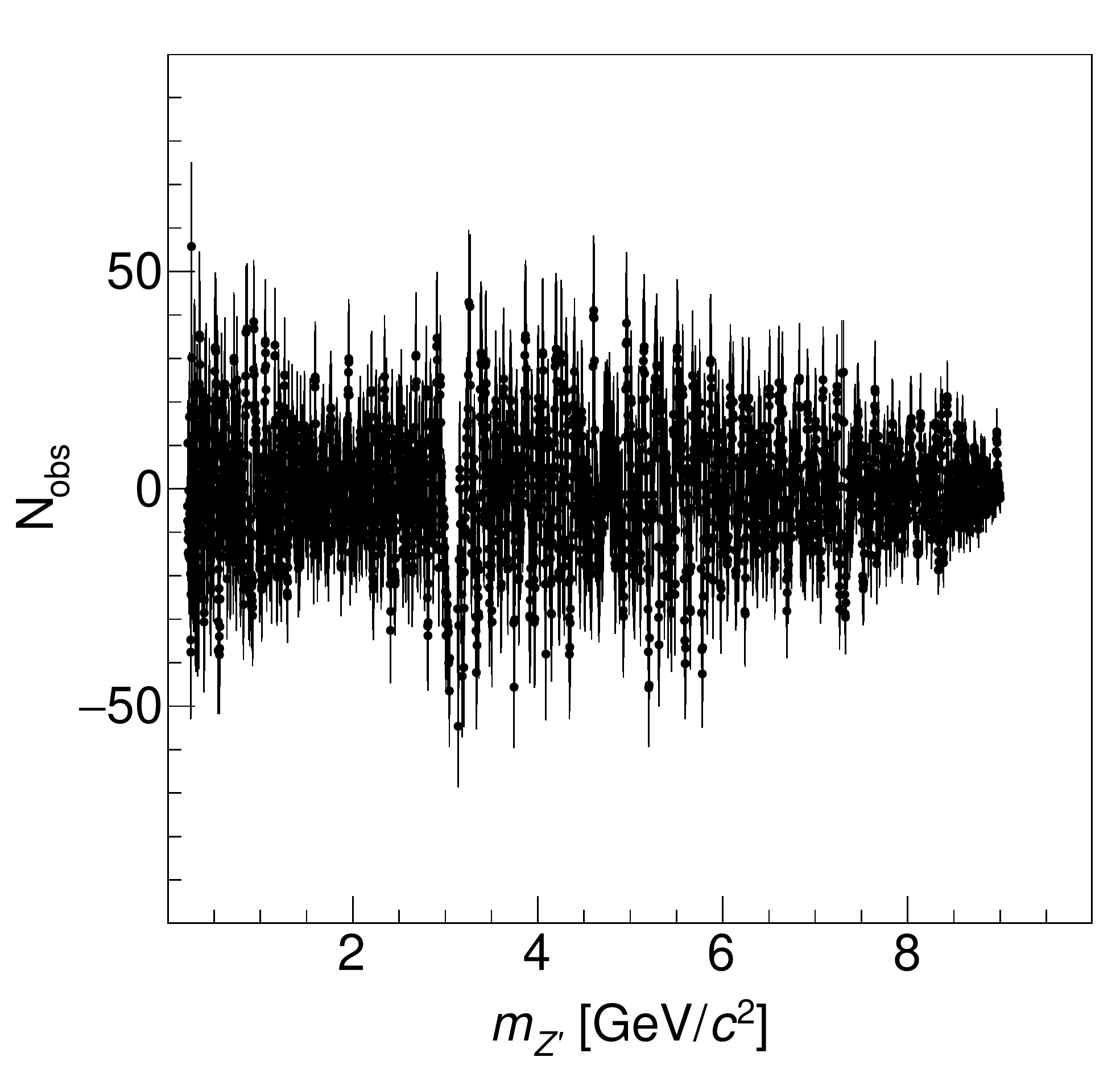}
  \caption{\label{fig::isryield}
Number of observed signal events ($N_\mathrm{obs}$) from the fit for the different $Z'$ mass hypotheses.
}
\end{figure}

The results are shown in Fig. \ref{fig::isryield}.
Using the calculated detection efficiency as shown in
Fig.~\ref{fig::deteff}, the branching fraction from Eq.\eqref{eq:zpmumu} and
the Belle luminosity $(\mathcal{L})$ of 643 fb$^{-1}$, the 90\% upper limit
on the Born $e^+ e^- \rightarrow Z' \mu^+ \mu^-$ cross-section is
obtained using:

  \begin{equation}\label{eq:bornsig}
    \sigma_\mathrm{B} = \frac{N_\mathrm{obs}}{\mathcal{L}\mathcal{B}\epsilon_{\mathrm{ISR}}(1 + \delta)|1-\Pi|^2}
   \end{equation}  
  where $N_\mathrm{obs}$, $\epsilon_{\mathrm{ISR}}$, $\mathcal{B}$, $(1 +
  \delta)$ and $|1-\Pi(s)|^2$ are the upper limit on the yield extracted from
  the data scan as shown in Fig.~\ref{fig::isryield}, the ISR signal MC sample based
  detection efficiency, the branching fraction from Eq.(\ref{eq:zpmumu}), the
  ISR correction factor, and the vacuum polarization factor, respectively.

 In order to test the ISR and the vaccum polarization effects, we check the ratio between the number of observed signal $N^{S}_{\mathrm{obs}}$ and the number of
  simulated signal events $N^{S}_\mathrm{MC}$ can be written as:
\begin{align}\label{eq:ratio}
  \frac{N_\mathrm{obs}^{S}}{N_\mathrm{MC}^{S}} = \frac{\sigma_V}{\sigma_B} \times \frac{\epsilon_\mathrm{ISR}}{\epsilon_\mathrm{non-ISR}}
\end{align}
where $\epsilon_\mathrm{ISR}(\epsilon_\mathrm{non-ISR})$ is the detection
efficiency obtained by the ISR (non-ISR) signal MC.
Since the cross-section with the ISR and vacuum polarization corrections ($\sigma_V$)
is related to the Born cross section by $\sigma_V = (1+\delta)|1-\Pi|^2 \times
\sigma_B$~\cite{Gribanov:2021amv} the ratio, Eq.(\ref{eq:ratio}), becomes:
\begin{align}
  \frac{N_\mathrm{obs}^S}{N_{\mathrm{MC}}^{S}}  =  (1+\delta)|1-\Pi|^2 \times \frac{\epsilon_{\mathrm{ISR}}}{\epsilon_{\mathrm{non-ISR}}} \nonumber \\
\end{align}

As the ISR and vacuum polarization corrections are common for the signal and the
$e^+ e^- \to \mu^+ \mu^- \mu^+ \mu^- (4\mu)$ background process, one can expect
that the ratio, Eq.(\ref{eq:ratio}), is the same for the signal and the $4\mu$
background:  $\frac{N_\mathrm{obs}^{S}}{N_\mathrm{MC}^{S}} =
\frac{N_\mathrm{obs}^{4\mu}}{N_\mathrm{MC}^{4\mu}}$.

Checking the consistency of the efficiency and ISR correction factors can be
carried out by the $4\mu$ MC background and data.
From Fig.~\ref{fig::redmasseff}, we observe the ratio between data and the MC
expectation for the $4\mu$ process to be:
$\frac{N^{S}_\mathrm{obs}}{N^{4\mu}_\mathrm{MC}} = 0.700$. This value is
compatible with the product of the ratio of the detection efficiencies
$\left(  \frac{\epsilon_\mathrm{ISR}}{\epsilon_\mathrm{non-ISR}} = 0.741\right)$
and the ISR
factor multiplied by the vacuum polarization ($(1 + \delta)(1 - \Pi)^2 = 0.945$).

\begin{figure}[h!]
  \centering
  \includegraphics[scale=0.4]{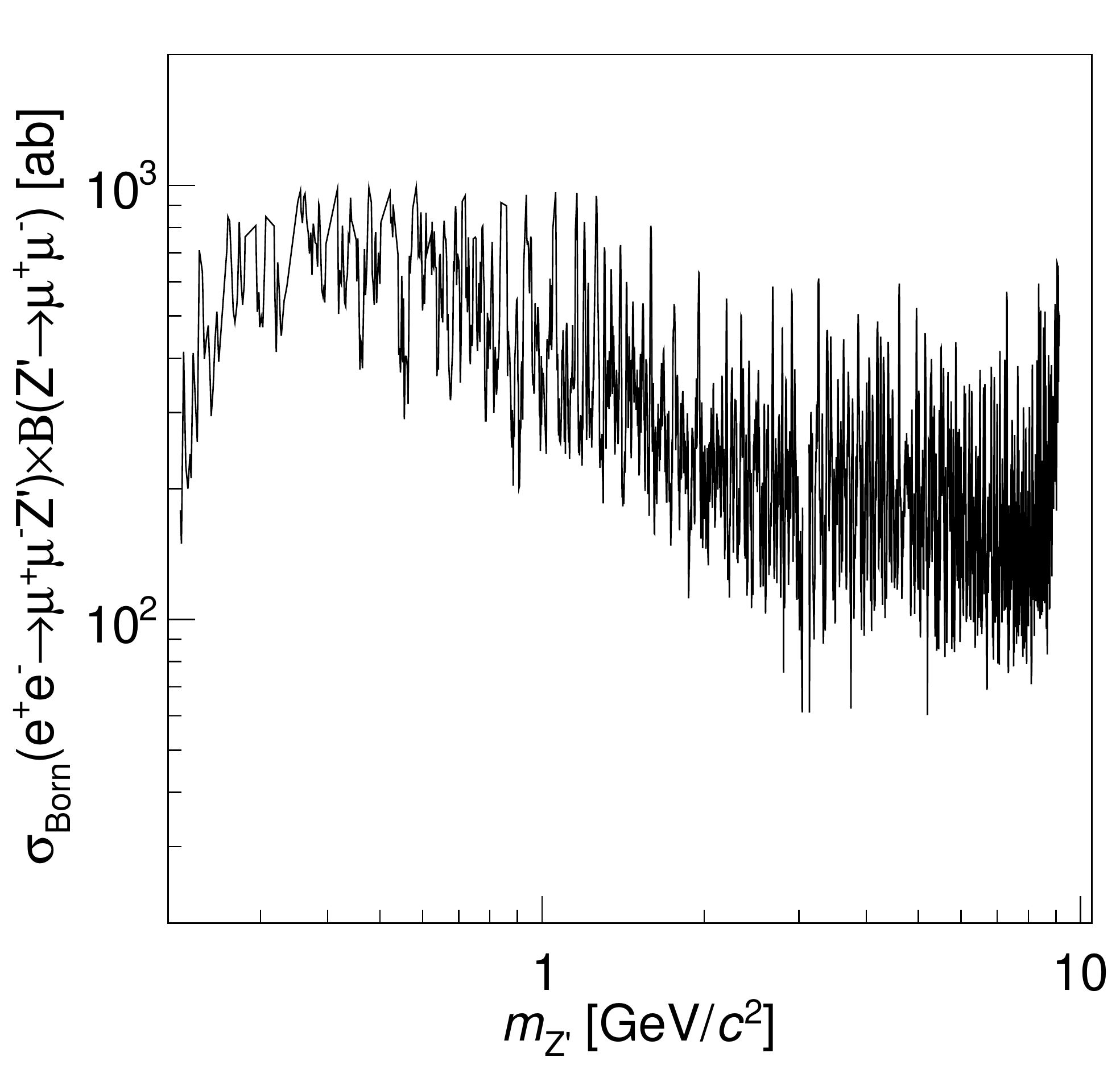}
  \caption{\label{fig::ninetylimxs} 90\% C.L. upper limit on the Born cross section
    for $e^+e^- \rightarrow \mu^+ \mu^- (Z' \rightarrow \mu^+ \mu^-)$
    as a function of $m_{Z'}$.
  }
  \end{figure}

The  90\% C.L. upper limits on Born cross-section as a function of $m_{Z'}$ are calculated and shown in Fig.~\ref{fig::ninetylimxs}.

\subsection{Limits on the Coupling Strength $g'$}

With a Born theoretical cross-section $\sigma_{\rm th}(\sqrt{s})$, for a given
$m_{Z'}$, at $\sqrt{s}$ and the coupling $g'$, the expected number of signal
events for data samples used in this analysis is given as:

\begin{align} \label{eq:cmsmerge}
N_\mathrm{exp} & = & \nonumber \\
  \qquad &  g'^2 \varepsilon \mathcal{B} \left(  \sigma^{\Upsilon(4S)}_{\mathrm{th}}(m_{Z'}) \mathcal{L}^{\Upsilon(4S)} + \sigma^{\Upsilon(3S)}_{\mathrm{th}} \mathcal{L}^{\Upsilon(3S)}  + \dots \right).
  \end{align}

\begin{figure}[h!]
  \includegraphics[scale=0.4]{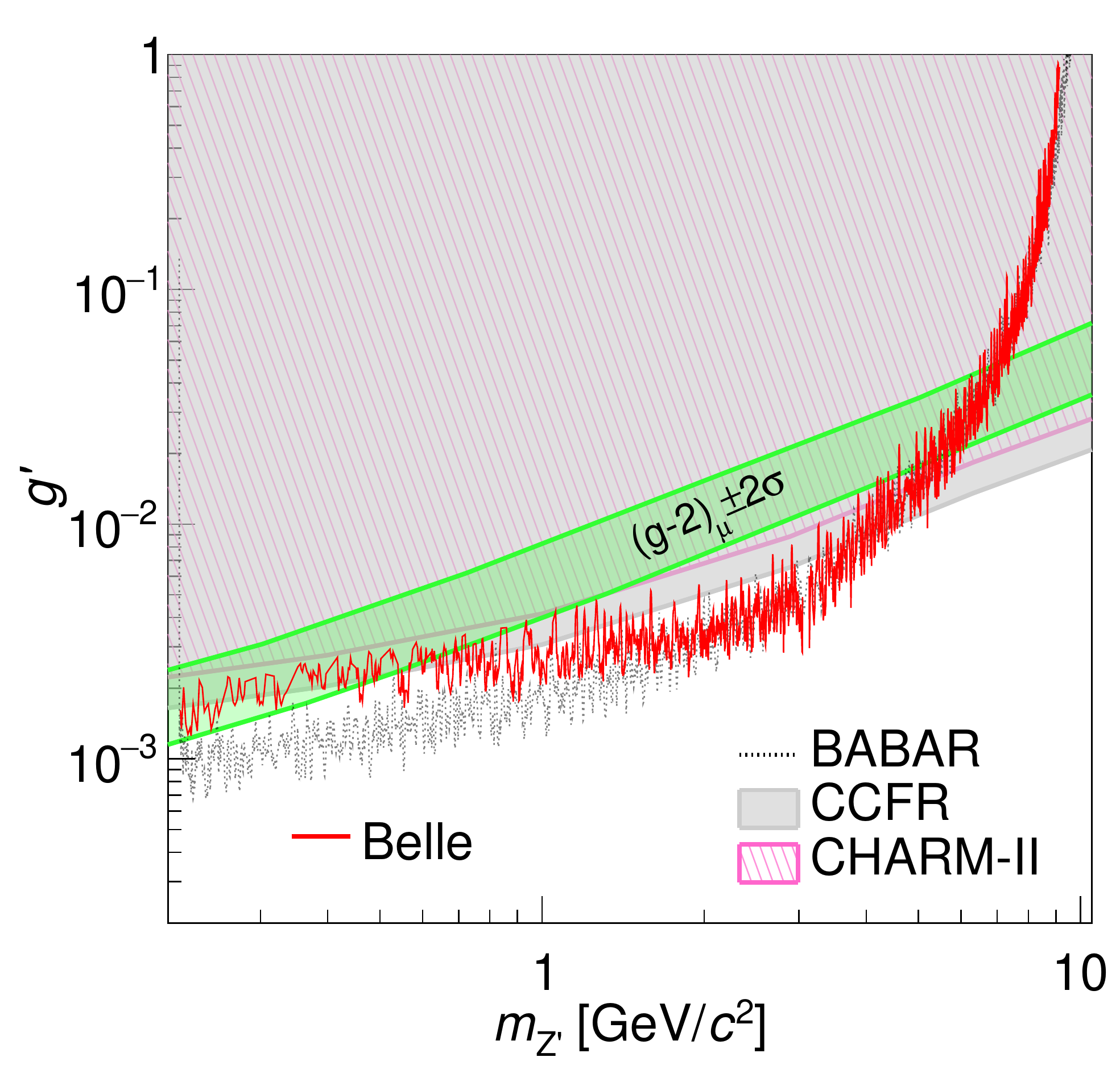}
  \caption{\label{fig::gpexpec}90\% C.L. upper limit on $g'$ as a function of
  $m_{Z'}$. The red solid the Belle result. The black dotted line represents the
  BABAR results
  ~\cite{babar2016zp}, the light gray shaded area is the result of CCFR, and the
  light purple hashed area the result of CHARM-II over the $Z'$ parameter space
  ~\cite{neutrinotrident, CCFR:1991lpl}, and the green region indicates the
  values of the $Z'$ coupling needed to explain $(g-2)_\mu$ suggested by the Muon $g-2$
  collaboration~\cite{gm2_2006, gm2_precision}.}
\end{figure}

  With Eq.\eqref{eq:cmsmerge}, the 90\% C.L. upper limit on $g'$ corresponding to
  $N_{\mathrm{exp}} = N_{\mathrm{obs}}$, is calculated and shown in
  Fig.~\ref{fig::gpexpec}.
  The result excludes most of the $Z'$ parameter space that could be related to
  the updated $(g-2)_\mu$ region, from the Muon $(g-2)$ experiment
  ~\cite{gm2_2006, gm2_precision}.
  Also shown in Fig.~\ref{fig::gpexpec} are comparisons with the CHARM-II experiment, the first measurement
  of the neutrino trident production ~\cite{CharmII}, the reinterpretation of the
  Columbia-Chicago-Fermilab-Rochester (CCFR) results ~\cite{neutrinotrident, CCFR:1991lpl} and the
first $Z' \rightarrow \mu^+ \mu^-$ search done by BABAR ~\cite{babar2016zp}.
  
\section{Conclusion}

In summary, we report a search for a new gauge boson $Z'$ in the
$L_\mu - L_\tau$ model with the on-shell production of $e^+ e^- \rightarrow Z'
\mu^+ \mu^-$, followed by $Z' \rightarrow \mu^+ \mu^-$. This is the first search
with the ISR effect directly included in the MC signal sample, while previous
searches did a data-driven estimation of the ISR effect. Since no significant excess is observed, the upper
limit on the coupling is set and the $Z'$ parameter space constraint is
improved.

This result specifically improves the previous $g'$ upper limit
between 2 and 8.4 GeV$/c^2$.

The $Z'$ mass region lighter than the dimuon threshold, does
not have any constraints but in the future,
 Belle II will be able to perform a more stringent test for the region \cite{zpinv, kaneta2016, araki2017}.

\section{Acknowledgments}

We thank B.~Shuve for providing the models for
\texttt{MadGraph5} and the branching fractions for $Z'$.
Our gratitude goes to K.~Mawatari for showing us the limitations of
\texttt{MadGraph5} when simulating ISR events and to J.~Reuter for
explaining how to use \texttt{WHIZARD} for generating ISR $Z'$ events.
We also thank T.~Shimomura for the enlightening discussions about $Z'$.

T. C. is supported by the Japan Society for the Promotion of Science (JSPS)
Grant No. 20H05858 and A. I. is supported by Grant No. 16H02176 and 22H00144.
We thank the KEKB group for the excellent operation of the
accelerator; the KEK cryogenics group for the efficient
operation of the solenoid; and the KEK computer group, and the Pacific Northwest National
Laboratory (PNNL) Environmental Molecular Sciences Laboratory (EMSL)
computing group for strong computing support; and the National
Institute of Informatics, and Science Information NETwork 5 (SINET5) for
valuable network support.  We acknowledge support from
the Ministry of Education, Culture, Sports, Science, and
Technology (MEXT) of Japan, the JSPS including Grant No. 20H05850, and the Tau-Lepton Physics 
Research Center of Nagoya University; 
the Australian Research Council including grants
DP180102629, 
DP170102389, 
DP170102204, 
DP150103061, 
FT130100303; 
Austrian Federal Ministry of Education, Science and Research (FWF) and
FWF Austrian Science Fund No.~P~31361-N36;
the National Natural Science Foundation of China under Contracts
No.~11435013,  
No.~11475187,  
No.~11521505,  
No.~11575017,  
No.~11675166,  
No.~11705209;  
Key Research Program of Frontier Sciences, Chinese Academy of Sciences (CAS), Grant No.~QYZDJ-SSW-SLH011; 
the  CAS Center for Excellence in Particle Physics (CCEPP); 
the Shanghai Pujiang Program under Grant No.~18PJ1401000;  
the Shanghai Science and Technology Committee (STCSM) under Grant No.~19ZR1403000; 
the Ministry of Education, Youth and Sports of the Czech
Republic under Contract No.~LTT17020;
Horizon 2020 ERC Advanced Grant No.~884719 and ERC Starting Grant No.~947006 ``InterLeptons'' (European Union);
the Carl Zeiss Foundation, the Deutsche Forschungsgemeinschaft, the
Excellence Cluster Universe, and the VolkswagenStiftung;
the Department of Atomic Energy (Project Identification No. RTI 4002) and the Department of Science and Technology of India; 
the Istituto Nazionale di Fisica Nucleare of Italy; 
National Research Foundation (NRF) of Korea Grant
Nos.~2016R1\-D1A1B\-01010135, 2016R1\-D1A1B\-02012900, 2018R1\-A2B\-3003643,
2018R1\-A6A1A\-06024970, 2018R1\-D1A1B\-07047294, 2019K1\-A3A7A\-09033840,
2019R1\-I1A3A\-01058933;
Radiation Science Research Institute, Foreign Large-size Research Facility Application Supporting project, the Global Science Experimental Data Hub Center of the Korea Institute of Science and Technology Information and KREONET/GLORIAD;
the Polish Ministry of Science and Higher Education and 
the National Science Center;
the Ministry of Science and Higher Education of the Russian Federation, Agreement 14.W03.31.0026, 
and the HSE University Basic Research Program, Moscow; 
University of Tabuk research grants
S-1440-0321, S-0256-1438, and S-0280-1439 (Saudi Arabia);
the Slovenian Research Agency Grant Nos. J1-9124 and P1-0135;
Ikerbasque, Basque Foundation for Science, Spain;
the Swiss National Science Foundation; 
the Ministry of Education and the Ministry of Science and Technology of Taiwan;
and the United States Department of Energy and the National Science Foundation.

\bibliographystyle{apsrev}
\bibliography{zparxiv.bib}

\end{document}